\documentclass[a4paper]{jpconf}
\usepackage{graphicx,float}
\usepackage{citesort}

\begin{document}
\title{Realistic shell-model calculations for neutron-rich calcium isotopes}

\author{L Coraggio$^1$, A Covello$^{1,2}$, A Gargano$^1$ and
  N Itaco$^{1,2}$}

\address{$^1$Istituto Nazionale di Fisica Nucleare and $^2$
  Dipartimento di Scienze Fisiche, Universit\`a di Napoli Federico II, \\
Complesso Universitario di Monte  S. Angelo, Via Cintia - I-80126 Napoli,
Italy}

\ead{luigi.coraggio@na.infn.it}

\begin{abstract}
We study the neutron-rich calcium isotopes performing shell-model
calculations with a realistic effective interaction.
This is derived from the CD-Bonn nucleon-nucleon potential
renormalized by way of the $V_{\rm low-k}$ approach, considering
$^{48}$Ca as an inert core and including the neutron
$0g_{9/2}$ orbital. We compare our results with experiment and with
the results of a previous study where $^{40}$Ca was assumed as a
closed core and the standard $1p0f$ model space was employed.
The calculated spectroscopic properties are in both cases in very good
agreement with the available experimental data and enable a
discussion on the role of the $g_{9/2}$ single-particle state in the
heavy-mass Ca isotopes.
\end{abstract}

\section{Introduction}
Heavy-mass calcium isotopes ($N>28$) have been recently the subject of
a renewed experimental and theoretical attention, since they lie far
from the stability valley, thus allowing to explore the evolution of the
shell structure when approaching the neutron drip line. 

This has resulted in a large number of experiments in this region
\cite{Liddick04ashort,Liddick04bshort,Dinca05short,Gade06short,Perrot06short,Rejmund07,Mantica08short,Fornal08short,Maierbeck09short,Bhattacharyya09short}
aiming to study the evolution of the single-particle (SP) orbitals.
On the theoretical side, to describe the spectroscopic properties of
these nuclei the shell model with various two-body effective
interactions has been extensively used in recent years.
The most frequently employed interactions are the KB3G \cite{Poves01},
FPD6 \cite{Richter91}, GXPF1 \cite{Honma02}, and a new version of the
latter dubbed GXPF1A \cite{Honma05}.
The above interactions have all been derived by modifying the
two-body matrix elements (TBME) of some realistic effective
interactions in order to obtain a better description of the
experimental data. 

During the last years, however, realistic shell-model effective
interactions without any empirical modifications have proved to
be able to describe with remarkable accuracy the spectroscopic
properties of nuclei in various mass regions
\cite{Coraggio06a,Coraggio07b,Coraggio09a,Coraggio09b}.

On these grounds, we have recently performed a shell-model study of
the Ca isotopic chain \cite{Coraggio09c} employing an effective
interaction derived from the CD-Bonn potential renormalized by way of
the $V_{\rm low-k}$ approach \cite{Bogner01,Bogner02}.
In that work we have chosen a model space spanned by the four neutron
SP levels of the $fp$ shell outside the doubly-closed $^{40}$Ca core.
This is a standard choice in most of the studies on neutron-rich Ca
isotopes, and has therefore allowed a direct comparison between our
results and those of other shell-model calculations.

In the present work, starting from the same $V_{\rm low-k}$ as in
Ref. \cite{Coraggio09c}, we have derived a shell-model effective
interaction for a model space spanned by the neutron  SP levels
$0f_{5/2}$, $1p_{3/2}$, $1p_{1/2}$, and $0g_{9/2}$ located above the
doubly-closed $^{48}$Ca core.
The aim of the present study is to compare the results of this
calculation with those of Ref. \cite{Coraggio09c}, so as to try to
clarify the role of the $0g_{9/2}$ orbital in heavy-mass calcium
isotopes.

The paper is organized as follows. 
In section 2 we give a brief outline of our calculations and some
details about the choice of the SP energies. 
Section 3 is devoted to the presentation of the results for calcium
isotopes beyond $^{48}$Ca up to $A$=56.
In section 4 we make some concluding remarks.

\section{Outline of calculations}
Our shell-model effective interaction has been derived within the
framework of perturbation theory \cite{Coraggio09a} starting from
the CD-Bonn $NN$ potential \cite{Machleidt01b}.
The high-momentum repulsive components of the bare $NN$ potential have
been renormalized by employing the so-called $V_{\rm low-k}$ approach
\cite{Bogner01,Bogner02}, so as to obtain a smooth potential that
preserves exactly the onshell properties of the original $NN$
potential.
Next, the TBME have been derived by way of the $\hat{Q}$-box plus
folded-diagram method \cite{Coraggio09a}, the $\hat{Q}$-box being a
collection of irreducible valence-linked Goldstone diagrams which we
have calculated through third order in $V_{\rm low-k}$.

The effective interaction $V_{\rm eff}$ can be written in an operator
form as 

\begin{equation}
V_{\rm eff} = \hat{Q} - \hat{Q'} \int \hat{Q} + \hat{Q'} \int \hat{Q} \int
\hat{Q} - \hat{Q'} \int \hat{Q} \int \hat{Q} \int \hat{Q} + ~...~~,
\end{equation}

\noindent
where the integral sign represents a generalized folding operation, 
and $\hat{Q'}$ is obtained from $\hat{Q}$ by removing terms of first
order in $V_{\rm low-k}$.
The folded-diagram series is summed up to all orders using the
Lee-Suzuki iteration method \cite{Suzuki80}.

We have derived our shell-model effective interaction for the
neutron-neutron, proton-proton, and proton-neutron channels using a
model space spanned by the four neutron SP levels $0f_{5/2}$,
$1p_{3/2}$, $1p_{1/2}$, $0g_{9/2}$, and by the sole proton SP level
$0f_{7/2}$, all of them located above the doubly-closed $^{48}$Ca core.
The complete list of the matrix elements is reported in table \ref{tbme}.
As already mentioned in the Introduction, in Ref. \cite{Coraggio09c}
the effective hamiltonian has been derived in a model space including
the four $pf$ orbitals outside doubly-closed $^{40}$Ca, that is the
standard choice of the model space in the study of calcium isotopes.

As regards the choice of the neutron SP energies, we refer to
\cite{Cottle08} for those of the $0f_{5/2}$, $1p_{3/2}$, $1p_{1/2}$
orbitals, obtained calculating the centroids of the observed
single-neutron strengths in $^{49}$Ca.
Since there is no experimental evidence of a SP $J^{\pi}=\frac{9}{2}^+$
state in $^{49}$Ca, the $0g_{9/2}$ SP energy has been fixed in order
to reproduce the energy of the SP $J^{\pi}=\frac{9}{2}^+$ state in
$^{57}$Ni \cite{Rudolph99}.
It is worth mentioning that this choice leads also to a good agreement
with experiment for the energies of $J^{\pi}=\frac{9}{2}^+$ states in
$^{51}$Ti and $^{53-57}$Cr.
The adopted SP energies are $\epsilon_{3/2}=-4.60$ MeV,
$\epsilon_{1/2}=-2.86$ MeV, $\epsilon_{5/2}=-1.20$ MeV, and
$\epsilon_{9/2}=-0.55$ MeV.

It should be pointed out that for protons the Coulomb force has been
explicitly added to the $V_{\rm low-k}$ before deriving $V_{\rm
  eff}$.

\section{Results}
In this section, we report the results of the present calculations for
the heavy-mass Ca isotopes from $^{50}$Ca to $^{56}$Ca (labelled as I)
and compare them with those of \cite{Coraggio09c} (labelled as II),
where $^{40}$Ca was assumed as a closed core.
The calculations have been performed by using the Oslo shell-model
code \cite{EngelandSMC}.

The calculated and experimental \cite{Rejmund07} energies of $^{50}$Ca
are reported in table \ref{50Catable}.
We see that both calculations reproduce quite well the observed
energies of positive-parity states.
The negative-parity states can be obviously obtained only in
calculation I, and the interpretation of their nature is strictly
connected with the corresponding states in the $^{48}$Ca energy
spectrum.

\begin{center}
\begin{table}[H]
\caption{\label{tbme} Proton-proton, neutron-neutron, and
  proton-neutron matrix elements (in MeV). They are antisymmetrized
  and normalized.}
\centering
\begin{tabular}{crrr|crrr}
\br
$n_a l_a j_a ~ n_b l_b j_b ~ n_c l_c j_c ~ n_d l_d j_d $ & $J$ &  $T_z$ &  TBME & $n_a l_a j_a ~ n_b l_b j_b ~ n_c l_c j_c ~ n_d l_d j_d $ & $J$ &  $T_z$ &  TBME \\
\mr
 $ 0f_{ 7/2}~ 0f_{ 7/2}~ 0f_{ 7/2}~ 0f_{ 7/2}$ &  0 &  1 & -1.867 & $ 0f_{ 5/2}~ 0g_{ 9/2}~ 0f_{ 5/2}~ 0g_{ 9/2}$ &  2 & -1 & -0.569 \\
 $ 0f_{ 7/2}~ 0f_{ 7/2}~ 0f_{ 7/2}~ 0f_{ 7/2}$ &  2 &  1 & -0.483 & $ 0f_{ 5/2}~ 0g_{ 9/2}~ 0f_{ 5/2}~ 0g_{ 9/2}$ &  3 & -1 & -0.436 \\
 $ 0f_{ 7/2}~ 0f_{ 7/2}~ 0f_{ 7/2}~ 0f_{ 7/2}$ &  4 &  1 &  0.160 & $ 0f_{ 5/2}~ 0g_{ 9/2}~ 1p_{ 3/2}~ 0g_{ 9/2}$ &  3 & -1 &  0.243 \\
 $ 0f_{ 7/2}~ 0f_{ 7/2}~ 0f_{ 7/2}~ 0f_{ 7/2}$ &  6 &  1 &  0.479 & $ 1p_{ 3/2}~ 0g_{ 9/2}~ 1p_{ 3/2}~ 0g_{ 9/2}$ &  3 & -1 & -0.827 \\
 $ 0f_{ 5/2}~ 0f_{ 5/2}~ 0f_{ 5/2}~ 0f_{ 5/2}$ &  0 & -1 & -0.841 & $ 0f_{ 5/2}~ 0g_{ 9/2}~ 0f_{ 5/2}~ 0g_{ 9/2}$ &  4 & -1 &  0.062 \\
 $ 0f_{ 5/2}~ 0f_{ 5/2}~ 1p_{ 3/2}~ 1p_{ 3/2}$ &  0 & -1 & -0.708 & $ 0f_{ 5/2}~ 0g_{ 9/2}~ 1p_{ 3/2}~ 0g_{ 9/2}$ &  4 & -1 &  0.170 \\
 $ 0f_{ 5/2}~ 0f_{ 5/2}~ 1p_{ 1/2}~ 1p_{ 1/2}$ &  0 & -1 & -0.396 & $ 0f_{ 5/2}~ 0g_{ 9/2}~ 1p_{ 1/2}~ 0g_{ 9/2}$ &  4 & -1 &  0.052 \\
 $ 0f_{ 5/2}~ 0f_{ 5/2}~ 0g_{ 9/2}~ 0g_{ 9/2}$ &  0 & -1 &  1.922 & $ 1p_{ 3/2}~ 0g_{ 9/2}~ 1p_{ 3/2}~ 0g_{ 9/2}$ &  4 & -1 &  0.060 \\
 $ 1p_{ 3/2}~ 1p_{ 3/2}~ 1p_{ 3/2}~ 1p_{ 3/2}$ &  0 & -1 & -0.802 & $ 1p_{ 3/2}~ 0g_{ 9/2}~ 1p_{ 1/2}~ 0g_{ 9/2}$ &  4 & -1 & -0.059 \\
 $ 1p_{ 3/2}~ 1p_{ 3/2}~ 1p_{ 1/2}~ 1p_{ 1/2}$ &  0 & -1 & -1.028 & $ 1p_{ 1/2}~ 0g_{ 9/2}~ 1p_{ 1/2}~ 0g_{ 9/2}$ &  4 & -1 &  0.112 \\
 $ 1p_{ 3/2}~ 1p_{ 3/2}~ 0g_{ 9/2}~ 0g_{ 9/2}$ &  0 & -1 &  1.305 & $ 0f_{ 5/2}~ 0g_{ 9/2}~ 0f_{ 5/2}~ 0g_{ 9/2}$ &  5 & -1 & -0.178 \\
 $ 1p_{ 1/2}~ 1p_{ 1/2}~ 1p_{ 1/2}~ 1p_{ 1/2}$ &  0 & -1 & -0.161 & $ 0f_{ 5/2}~ 0g_{ 9/2}~ 1p_{ 3/2}~ 0g_{ 9/2}$ &  5 & -1 &  0.196 \\
 $ 1p_{ 1/2}~ 1p_{ 1/2}~ 0g_{ 9/2}~ 0g_{ 9/2}$ &  0 & -1 &  0.980 & $ 0f_{ 5/2}~ 0g_{ 9/2}~ 1p_{ 1/2}~ 0g_{ 9/2}$ &  5 & -1 & -0.275 \\
 $ 0g_{ 9/2}~ 0g_{ 9/2}~ 0g_{ 9/2}~ 0g_{ 9/2}$ &  0 & -1 & -1.067 & $ 1p_{ 3/2}~ 0g_{ 9/2}~ 1p_{ 3/2}~ 0g_{ 9/2}$ &  5 & -1 & -0.191 \\
 $ 0f_{ 5/2}~ 1p_{ 3/2}~ 0f_{ 5/2}~ 1p_{ 3/2}$ &  1 & -1 &  0.138 & $ 1p_{ 3/2}~ 0g_{ 9/2}~ 1p_{ 1/2}~ 0g_{ 9/2}$ &  5 & -1 &  0.487 \\
 $ 0f_{ 5/2}~ 1p_{ 3/2}~ 1p_{ 3/2}~ 1p_{ 1/2}$ &  1 & -1 & -0.061 & $ 1p_{ 1/2}~ 0g_{ 9/2}~ 1p_{ 1/2}~ 0g_{ 9/2}$ &  5 & -1 & -0.401 \\
 $ 1p_{ 3/2}~ 1p_{ 1/2}~ 1p_{ 3/2}~ 1p_{ 1/2}$ &  1 & -1 &  0.244 & $ 0f_{ 5/2}~ 0g_{ 9/2}~ 0f_{ 5/2}~ 0g_{ 9/2}$ &  6 & -1 &  0.153 \\
 $ 0f_{ 5/2}~ 0f_{ 5/2}~ 0f_{ 5/2}~ 0f_{ 5/2}$ &  2 & -1 & -0.207 & $ 0f_{ 5/2}~ 0g_{ 9/2}~ 1p_{ 3/2}~ 0g_{ 9/2}$ &  6 & -1 &  0.194 \\
 $ 0f_{ 5/2}~ 0f_{ 5/2}~ 0f_{ 5/2}~ 1p_{ 3/2}$ &  2 & -1 &  0.089 & $ 1p_{ 3/2}~ 0g_{ 9/2}~ 1p_{ 3/2}~ 0g_{ 9/2}$ &  6 & -1 &  0.400 \\
 $ 0f_{ 5/2}~ 0f_{ 5/2}~ 0f_{ 5/2}~ 1p_{ 1/2}$ &  2 & -1 & -0.232 & $ 0f_{ 5/2}~ 0g_{ 9/2}~ 0f_{ 5/2}~ 0g_{ 9/2}$ &  7 & -1 & -0.945 \\
 $ 0f_{ 5/2}~ 0f_{ 5/2}~ 1p_{ 3/2}~ 1p_{ 3/2}$ &  2 & -1 & -0.126 & $ 0f_{ 7/2}~ 0f_{ 5/2}~ 0f_{ 7/2}~ 0f_{ 5/2}$ &  1 &  0 & -2.430 \\
 $ 0f_{ 5/2}~ 0f_{ 5/2}~ 1p_{ 3/2}~ 1p_{ 1/2}$ &  2 & -1 & -0.285 & $ 0f_{ 7/2}~ 0f_{ 5/2}~ 0f_{ 7/2}~ 0f_{ 5/2}$ &  2 &  0 & -1.562 \\
 $ 0f_{ 5/2}~ 0f_{ 5/2}~ 0g_{ 9/2}~ 0g_{ 9/2}$ &  2 & -1 &  0.266 & $ 0f_{ 7/2}~ 0f_{ 5/2}~ 0f_{ 7/2}~ 1p_{ 3/2}$ &  2 &  0 & -0.638 \\
 $ 0f_{ 5/2}~ 1p_{ 3/2}~ 0f_{ 5/2}~ 1p_{ 3/2}$ &  2 & -1 &  0.219 & $ 0f_{ 7/2}~ 1p_{ 3/2}~ 0f_{ 7/2}~ 1p_{ 3/2}$ &  2 &  0 & -0.830 \\
 $ 0f_{ 5/2}~ 1p_{ 3/2}~ 0f_{ 5/2}~ 1p_{ 1/2}$ &  2 & -1 &  0.353 & $ 0f_{ 7/2}~ 0f_{ 5/2}~ 0f_{ 7/2}~ 0f_{ 5/2}$ &  3 &  0 & -0.422 \\
 $ 0f_{ 5/2}~ 1p_{ 3/2}~ 1p_{ 3/2}~ 1p_{ 3/2}$ &  2 & -1 &  0.133 & $ 0f_{ 7/2}~ 0f_{ 5/2}~ 0f_{ 7/2}~ 1p_{ 3/2}$ &  3 &  0 &  0.169 \\
 $ 0f_{ 5/2}~ 1p_{ 3/2}~ 1p_{ 3/2}~ 1p_{ 1/2}$ &  2 & -1 &  0.221 & $ 0f_{ 7/2}~ 0f_{ 5/2}~ 0f_{ 7/2}~ 1p_{ 1/2}$ &  3 &  0 & -0.295 \\
 $ 0f_{ 5/2}~ 1p_{ 3/2}~ 0g_{ 9/2}~ 0g_{ 9/2}$ &  2 & -1 & -0.554 & $ 0f_{ 7/2}~ 1p_{ 3/2}~ 0f_{ 7/2}~ 1p_{ 3/2}$ &  3 &  0 & -0.369 \\
 $ 0f_{ 5/2}~ 1p_{ 1/2}~ 0f_{ 5/2}~ 1p_{ 1/2}$ &  2 & -1 & -0.197 & $ 0f_{ 7/2}~ 1p_{ 3/2}~ 0f_{ 7/2}~ 1p_{ 1/2}$ &  3 &  0 &  0.790 \\
 $ 0f_{ 5/2}~ 1p_{ 1/2}~ 1p_{ 3/2}~ 1p_{ 3/2}$ &  2 & -1 & -0.127 & $ 0f_{ 7/2}~ 1p_{ 1/2}~ 0f_{ 7/2}~ 1p_{ 1/2}$ &  3 &  0 & -0.678 \\
 $ 0f_{ 5/2}~ 1p_{ 1/2}~ 1p_{ 3/2}~ 1p_{ 1/2}$ &  2 & -1 & -0.386 & $ 0f_{ 7/2}~ 0f_{ 5/2}~ 0f_{ 7/2}~ 0f_{ 5/2}$ &  4 &  0 & -0.891 \\
 $ 0f_{ 5/2}~ 1p_{ 1/2}~ 0g_{ 9/2}~ 0g_{ 9/2}$ &  2 & -1 &  0.684 & $ 0f_{ 7/2}~ 0f_{ 5/2}~ 0f_{ 7/2}~ 1p_{ 3/2}$ &  4 &  0 & -0.189 \\
 $ 1p_{ 3/2}~ 1p_{ 3/2}~ 1p_{ 3/2}~ 1p_{ 3/2}$ &  2 & -1 & -0.308 & $ 0f_{ 7/2}~ 0f_{ 5/2}~ 0f_{ 7/2}~ 1p_{ 1/2}$ &  4 &  0 & -0.546 \\
 $ 1p_{ 3/2}~ 1p_{ 3/2}~ 1p_{ 3/2}~ 1p_{ 1/2}$ &  2 & -1 & -0.534 & $ 0f_{ 7/2}~ 1p_{ 3/2}~ 0f_{ 7/2}~ 1p_{ 3/2}$ &  4 &  0 & -0.085 \\
 $ 1p_{ 3/2}~ 1p_{ 3/2}~ 0g_{ 9/2}~ 0g_{ 9/2}$ &  2 & -1 &  0.477 & $ 0f_{ 7/2}~ 1p_{ 3/2}~ 0f_{ 7/2}~ 1p_{ 1/2}$ &  4 &  0 & -0.344 \\
 $ 1p_{ 3/2}~ 1p_{ 1/2}~ 1p_{ 3/2}~ 1p_{ 1/2}$ &  2 & -1 & -0.452 & $ 0f_{ 7/2}~ 1p_{ 1/2}~ 0f_{ 7/2}~ 1p_{ 1/2}$ &  4 &  0 & -0.573 \\
 $ 1p_{ 3/2}~ 1p_{ 1/2}~ 0g_{ 9/2}~ 0g_{ 9/2}$ &  2 & -1 &  0.397 & $ 0f_{ 7/2}~ 0f_{ 5/2}~ 0f_{ 7/2}~ 0f_{ 5/2}$ &  5 &  0 &  0.157 \\
 $ 0g_{ 9/2}~ 0g_{ 9/2}~ 0g_{ 9/2}~ 0g_{ 9/2}$ &  2 & -1 & -0.771 & $ 0f_{ 7/2}~ 0f_{ 5/2}~ 0f_{ 7/2}~ 1p_{ 3/2}$ &  5 &  0 &  0.246 \\
 $ 0f_{ 5/2}~ 1p_{ 3/2}~ 0f_{ 5/2}~ 1p_{ 3/2}$ &  3 & -1 &  0.163 & $ 0f_{ 7/2}~ 1p_{ 3/2}~ 0f_{ 7/2}~ 1p_{ 3/2}$ &  5 &  0 & -1.027 \\
 $ 0f_{ 5/2}~ 1p_{ 3/2}~ 0f_{ 5/2}~ 1p_{ 1/2}$ &  3 & -1 &  0.044 & $ 0f_{ 7/2}~ 0f_{ 5/2}~ 0f_{ 7/2}~ 0f_{ 5/2}$ &  6 &  0 & -1.796 \\
 $ 0f_{ 5/2}~ 1p_{ 1/2}~ 0f_{ 5/2}~ 1p_{ 1/2}$ &  3 & -1 &  0.307 & $ 0f_{ 7/2}~ 0g_{ 9/2}~ 0f_{ 7/2}~ 0g_{ 9/2}$ &  1 &  0 & -1.138 \\
 $ 0f_{ 5/2}~ 0f_{ 5/2}~ 0f_{ 5/2}~ 0f_{ 5/2}$ &  4 & -1 &  0.198 & $ 0f_{ 7/2}~ 0g_{ 9/2}~ 0f_{ 7/2}~ 0g_{ 9/2}$ &  2 &  0 & -0.846 \\
 $ 0f_{ 5/2}~ 0f_{ 5/2}~ 0f_{ 5/2}~ 1p_{ 3/2}$ &  4 & -1 &  0.207 & $ 0f_{ 7/2}~ 0g_{ 9/2}~ 0f_{ 7/2}~ 0g_{ 9/2}$ &  3 &  0 & -0.288 \\
 $ 0f_{ 5/2}~ 0f_{ 5/2}~ 0g_{ 9/2}~ 0g_{ 9/2}$ &  4 & -1 &  0.228 & $ 0f_{ 7/2}~ 0g_{ 9/2}~ 0f_{ 7/2}~ 0g_{ 9/2}$ &  4 &  0 & -0.331 \\
 $ 0f_{ 5/2}~ 1p_{ 3/2}~ 0f_{ 5/2}~ 1p_{ 3/2}$ &  4 & -1 & -0.243 & $ 0f_{ 7/2}~ 0g_{ 9/2}~ 0f_{ 7/2}~ 0g_{ 9/2}$ &  5 &  0 & -0.066 \\
 $ 0f_{ 5/2}~ 1p_{ 3/2}~ 0g_{ 9/2}~ 0g_{ 9/2}$ &  4 & -1 & -0.613 & $ 0f_{ 7/2}~ 0g_{ 9/2}~ 0f_{ 7/2}~ 0g_{ 9/2}$ &  6 &  0 & -0.499 \\
 $ 0g_{ 9/2}~ 0g_{ 9/2}~ 0g_{ 9/2}~ 0g_{ 9/2}$ &  4 & -1 & -0.277 & $ 0f_{ 7/2}~ 0g_{ 9/2}~ 0f_{ 7/2}~ 0g_{ 9/2}$ &  7 &  0 &  0.016 \\
 $ 0g_{ 9/2}~ 0g_{ 9/2}~ 0g_{ 9/2}~ 0g_{ 9/2}$ &  6 & -1 & -0.116 & $ 0f_{ 7/2}~ 0g_{ 9/2}~ 0f_{ 7/2}~ 0g_{ 9/2}$ &  8 &  0 & -1.795 \\
 $ 0g_{ 9/2}~ 0g_{ 9/2}~ 0g_{ 9/2}~ 0g_{ 9/2}$ &  8 & -1 & -0.004 & ~ &  ~ &  ~ & ~ \\
\br
\end{tabular}
\end{table}
\end{center}

As a matter of fact, in $^{48}$Ca the $J^{\pi}=3^-_1,~4^-_1,~5^-_1$
states, which can be interpreted as $1p-1h$ $Z=20$ cross-shell
excitations, are located $\sim 0.5$ MeV higher in energy than the
corresponding ones in $^{50}$Ca.
This may be seen as an indication that, when increasing the number of
neutrons, the structure of the above negative-parity states changes
because of configuration mixing with states built on the neutron
$0g_{9/2}$ orbital.
Our results seem to indicate that the overlap with our model space is
larger for $J^{\pi}=5^-_1$ than for $J^{\pi}=3^-_1,~4^-_1$.
The experimental $J^{\pi}=2^+_3,~4^+_1$ states have no theoretical
counterpart in calculation I since their structure is dominated by the
$\nu (0f_{7/2})^7 (1p_{3/2})^3$ configuration
\cite{Coraggio09c,Rejmund07}.
Note that the results for positive-parity states with calculations I
and II are close to each other, thus confirming the soundness of the
perturbation theory of the shell-model effective interaction.
We have also calculated the $B(E2;2^+_1 \rightarrow 0^+_1)$ transition
rate in $^{50}$Ca employing effective operators obtained at third
order in perturbation theory, consistently with the derivation of
$H_{\rm eff}$ as in Ref. \cite{Coraggio09c}.
The calculated value is $7.1~{\rm e^2 fm^4}$ to be compared with the
experimental one $7.5 \pm 0.2~ {\rm e^2 fm^4}$
\cite{Valiente-Dobon09short}, while with calculation II we have
obtained $10.9~{\rm e^2 fm^4}$ \cite{Coraggio09c}.

\vspace{-0.8truecm}

\begin{center}
\begin{table}[H]
\caption{\label{50Catable} Calculated energy levels
  (in MeV) of $^{50}$Ca and $^{52}$Ca  compared with the corresponding
  experimental ones \cite{Rejmund07,nndc} up to 5.2 MeV excitation
  energy. The energies of the states whose parity and/or angular
  momentum are uncertain are reported in parenthesis.}
\centering
\begin{tabular}{ccccc|ccccc}
\br
 ~~ &   ~~ &  $^{50}$Ca & ~~~ & ~~~ & ~~ & ~~ &   ~~ &  $^{52}$Ca & ~~~ \\
$J^{\pi}$ & Expt. & Calc. I & Calc. II & ~ & ~ & $J^{\pi}$ & Expt. & Calc. I & Calc. II \\
\mr
$0^+$  & 0.000   & 0.000 & 0.000 & ~ & ~ & $0^+$  & 0.000   & 0.000 & 0.000 \\
$2^+$  & 1.026   & 0.939 & 0.953 & ~ & ~ & $2^+$  & 2.562   & 2.358 & 2.396 \\
$2^+$  & (3.004) & 2.856 & 2.941 & ~ & ~~ & $1^+$ & (3.150) & 3.205 & 3.135 \\
$1^+$  & (3.532) & 3.472 & 3.547 & ~ & ~ & $3^-$  & (3.992) & 4.702 & ~~~~  \\
$3^-$  & (3.998) & 4.698 & ~~~~  & ~ & ~ & ~~      & ~~     & ~~    & ~~ \\
$2^+$  & (4.036) & ~~~   & 3.582 & ~ & ~ & ~~      & ~~     & ~~    & ~~ \\
$0^+$  & (4.470) & 4.839 & 4.926 & ~ & ~ & ~~      & ~~     & ~~    & ~~ \\
$4^+$  & 4.515   & ~~~   & 4.567 & ~ & ~ & ~~      & ~~     & ~~    & ~~ \\
$4^-$  & (4.832) & 5.590 & ~~~~~ & ~ & ~ & ~~      & ~~     & ~~    & ~~ \\
$2^+$  & (4.870) & 4.957 & 5.231 & ~ & ~ & ~~      & ~~     & ~~    & ~~ \\
$5^-$  & (5.110) & 5.188 & ~~~~~ & ~ & ~ & ~~      & ~~     & ~~    & ~~ \\
\br
\end{tabular}
\end{table}
\end{center}

In table \ref{51Catable} we have reported the experimental and
calculated energy spectra of $^{51}$Ca.
We see that also for this nucleus the comparison between theory and
experiment is remarkably good, apart from the
$J^{\pi}=\frac{7}{2}^+_{\scriptscriptstyle 1}$ state.
The structure of this state is mainly obtained coupling the
$J^{\pi}=3^-_1$ state in $^{50}$Ca to a neutron in the $1p_{3/2}$
orbital, which explains the overestimation of the calculated energy by
$\sim 450$ keV.
In calculation I we find that the first excited
$J^{\pi}=\frac{9}{2}^+$ state is essentially a seniority $v=1$ state.
This lends further support to our choice of the $0g_{9/2}$ SP energy.
It should be mentioned that in Ref. \cite{Fornal08short} both
$J^{\pi}=\frac{7}{2}^+_{\scriptscriptstyle 1}$ and
$J^{\pi}=\frac{9}{2}^+_{\scriptscriptstyle 1}$ are interpreted as
$1p-1h$ $Z=20$ cross-shell excitations.

As regards the negative-parity states, the agreement between
calculations I and II is very good.
Note that the $J^{\pi}=\frac{7}{2}^-_{\scriptscriptstyle 1}$ state is
outside the model space of calculation I since its structure is mainly
made up by one neutron-hole in $^{48}$Ca core and 4 neutrons filling
the $1p_{3/2}$ orbital.

The nucleus $^{52}$Ca is the last one of this isotopic chain for which
some experimental excited states have been observed
\cite{Rejmund07,Gade06short,Perrot06short}.
The experimental energies of these states are reported in table
\ref{50Catable}  together with the corresponding calculated ones up to
5.2 MeV.
Both calculations I and II reproduce quite well the subshell closure
of the $1p_{3/2}$ orbital.

\vspace{-0.8truecm}
\begin{center}
\begin{table}[H]
\caption{\label{51Catable} Energy levels
  (in MeV) of $^{51}$Ca and $^{53}$Ca. The calculated $^{51}$Ca energy
  levels are compared with the corresponding experimental ones
  \cite{Rejmund07,Fornal08short}. The energies of the states whose parity
  and/or angular momentum are uncertain are reported in
  parenthesis. The $^{53}$Ca energy levels are reported up to 4.0 MeV
  excitation energy.}
\centering
\begin{tabular}{ccccc|cccc}
\br
 ~~ &   ~~ &  $^{51}$Ca & ~~~ & ~~~ & ~~ & ~~ &  $^{53}$Ca & ~~~ \\
$J^{\pi}$ & Expt. & Calc. I & Calc. II & ~ & ~ & $J^{\pi}$ & Calc. I & Calc. II \\
\mr
$\frac{3}{2}^-$ & (0.000) & 0.000 & 0.000 & ~ & ~ & $\frac{1}{2}^-$ & 0.000 & 0.000 \\
$\frac{1}{2}^-$ & (1.721) & 1.753 & 1.586 & ~ & ~ & $\frac{5}{2}^-$ & 2.112 & 2.039 \\
$\frac{5}{2}^-$ & (2.379) & 2.190 & 2.269 & ~ & ~ & $\frac{3}{2}^-$ & 2.169 & 2.402
\\
$\frac{3}{2}^-$ & (2.937) & 2.829 & 2.932 & ~ & ~ & $\frac{9}{2}^+$ & 2.604 & ~~\\
$\frac{7}{2}^-$ & (3.437) & ~~~~~ & 3.429 & ~ & ~ & ~~      & ~~    & ~~ \\
$\frac{5}{2}^-$ & (3.479) & 3.564 & 3.552 & ~ & ~ & ~~      & ~~    & ~~ \\
$\frac{7}{2}^+$ & (3.845) & 4.309 & ~~~~~ & ~ & ~ & ~~      & ~~    & ~~ \\
$\frac{9}{2}^+$ & (4.155) & 4.126 & ~~~~~ & ~ & ~ & ~~      & ~~    & ~~ \\
$\frac{9}{2}^-$ & (4.322) & 4.281 & 4.435 & ~ & ~ & ~~      & ~~    & ~~ \\
\br
\end{tabular}
\end{table}
\end{center}

\begin{figure}[H]
\includegraphics[width=22pc]{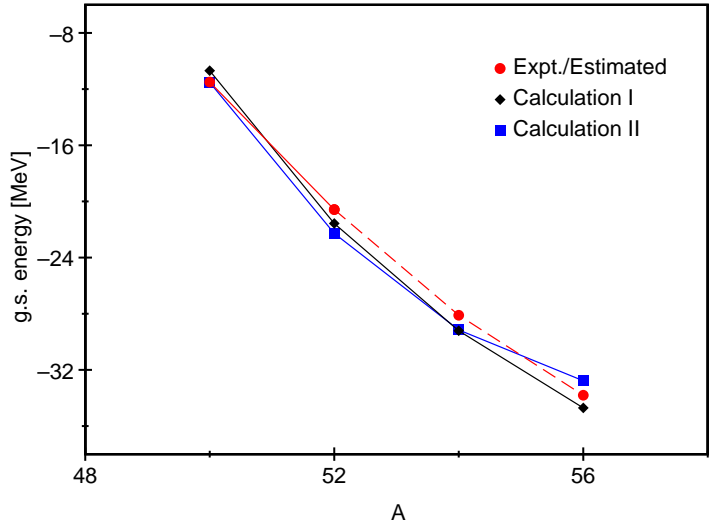}\hspace{2pc}%
\begin{minipage}[b]{14pc}\caption{\label{Cagse} Ground-state energies
    for calcium isotopes from $A=50$ to 56. See text for details.}
\end{minipage}
\end{figure}

\vspace{-0.4truecm}
For the heavier Ca isotopes, $^{53,54,55}$Ca, only the ground states
have been identified with spin and parity assignments
\cite{Mantica08short}.
In tables \ref{51Catable} and \ref{54Catable} we have reported our
predicted spectra for these three nuclei and for $^{56}$Ca up to 4.0,
4.5, 2.4, and 3.0 MeV excitation energy, respectively.
From the inspection of the above tables, it is clear that the
disagreement between calculations I and II increases with mass number,
since the role of $0g_{9/2}$ orbital becomes more and more relevant
when adding valence neutrons.
This has also repercussions on the two-neutron separation energy
$S_{2n}$ for $^{56}$Ca which is about 2 MeV larger in calculation I,
as can be inferred from the inspection of figure \ref{Cagse} where the
calculated ground-state energies of the even-mass isotopes relative to
$^{48}$Ca are compared with the measured ($A=50,~52$) and estimated
ones ($A=54,~56$) \cite{Audi03}.

\vspace{-0.4truecm}
\begin{center}
\begin{table}[H]
\caption{\label{54Catable} Theoretical energy levels (in MeV) of
    $^{54}$Ca, $^{55}$Ca, and $^{56}$Ca. See text for details.}
\centering
\begin{tabular}{cccc|ccccc|cccc}
\br
 ~~ &  $^{54}$Ca & ~~~ & ~~~ & ~~ & ~~ &  $^{55}$Ca & ~~~ & ~~~ & ~~ & ~~ &  $^{56}$Ca & ~~~ \\
$J^{\pi}$ & Calc. I & Calc. II & ~ & ~ & $J^{\pi}$ & Calc. I & Calc. II & ~ & ~ & $J^{\pi}$ & Calc. I & Calc. II \\
\mr
$0^+$  & 0.000 & 0.000 & ~ & ~ & $\frac{5}{2}^-$  & 0.000 & 0.000 & ~ & ~ & $0^+$  & 0.000 & 0.000 \\
$2^+$  & 2.587 & 2.061 & ~ & ~ & $\frac{9}{2}^+$  & 0.208 & ~~~~~ & ~ & ~ & $2^+$  & 2.268 & 0.963 \\
$0^+$  & 2.783 & 3.138 & ~ & ~ & $\frac{1}{2}^-$  & 0.237 & 1.104 & ~ & ~ & $7^-$  & 2.514 & ~~~~~ \\
$5^-$  & 3.116 & ~~    & ~ & ~ & $\frac{3}{2}^-$  & 1.924 & 1.541 & ~ & ~ & $2^-$  & 2.745 & ~~~~~ \\
$3^+$  & 3.157 & 2.748 & ~ & ~ & $\frac{5}{2}^+$  & 2.131 & ~~~~~ & ~ & ~ & $3^-$  & 2.825 & ~~~~~ \\
$4^-$  & 3.430 & ~~~   & ~ & ~ & $\frac{13}{2}^+$ & 2.204 & ~~~~~ & ~ & ~ & $5^-$  & 2.960 & ~~~~~ \\
$2^+$  & 4.517 & 4.172 & ~ & ~ & $\frac{7}{2}^-$  & 2.325 & 1.931 & ~ & ~ & $4^+$  & 2.961 & 1.487 \\
~~~~~~ & ~~~~~ & ~~~~~ & ~ & ~ & $\frac{7}{2}^+$  & 2.368 & ~~~~~ & ~ & ~ & $2^+$  & 2.982 & 2.224 \\
~~~~~~ & ~~~~~ & ~~~~~ & ~ & ~ & $\frac{5}{2}^-$  & 2.381 & 2.390 & ~ & ~ & ~~~  & ~~~~ & ~~~~ \\
\br
\end{tabular}
\end{table}
\end{center}

\section{Concluding remarks}
In summary, we have given here a shell-model description of heavy
calcium isotopes assuming $^{48}$Ca as a closed core and
including the neutron $0g_{9/2}$ orbital.
This has been done by renormalizing the $NN$ potential CD-Bonn by way
of the $V_{\rm low-k}$ approach and then deriving the shell-model
hamiltonian within the framework of perturbation theory.
We have compared the results of the present study with those of the previous 
one \cite{Coraggio09c} which started from doubly-closed $^{40}$Ca and
employed the $1p0f$ model space.

The results of both calculations are in good agreement with the
available experimental data, and do not differ significantly up to
$^{53}$Ca.
Actually, from $^{54}$Ca on the differences between the energy
spectra calculated within the two different model spaces increase with
the number of valence neutrons.
Clearly, new experimental data are needed to establish the real
role of the neutron $0g_{9/2}$ orbital in the description of the
structure of nuclei north-east of $^{48}$Ca core.

\section*{References}
\bibliography{biblio}

\end{document}